\documentclass[10pt,aps,prl,twocolumn,showpacs,preprintnumbers,amsmath,amssymb,superscriptaddress,longbibliography
]{revtex4-2}
\pdfoutput=1

\usepackage{amsfonts,amsmath,amssymb}
\usepackage{epsf}
\usepackage{braket}
\usepackage{here}
\usepackage{mathtools}
\usepackage{physics}
\usepackage{graphicx,hyperref,color}
\usepackage[dvipsnames]{xcolor}
\usepackage{url}
\usepackage{comment}
\usepackage{tikz}
\usepackage{circuitikz}
\usepackage{lipsum}
\usepackage[capitalize]{cleveref}
\usepackage{multibib}
    \newcites{SM}{Supplementary References}
\usepackage{appendix}
\usepackage{chngcntr}
\usepackage{etoolbox}

\hypersetup{
    unicode=false,          
    pdftoolbar=true,        
    pdfmenubar=true,        
    linktocpage=true,       
    pdffitwindow=false,     
    pdfstartview={FitH},    
    pdfnewwindow=true,      
    colorlinks=true,       
    linkcolor=blue,          
    citecolor=blue,        
    filecolor=blue,      
    urlcolor=blue           
}

\newcommand{\be}{\begin{equation}}
\newcommand{\ba}{\begin{eqnarray}}
\newcommand{\ea}{\end{eqnarray}}
\newcommand{\ee}{\end{equation}}

\newcommand{\bea}{\begin{eqnarray}}
\newcommand{\eea}{\end{eqnarray}}
\newcommand{\bes}{\begin{equation*}}
\newcommand{\beas}{\begin{eqnarray*}}
\newcommand{\eeas}{\end{eqnarray*}}
\newcommand{\ees}{\end{equation*}}

\newcommand{\ep}{\epsilon}

\newcommand{\ie}{\textit{i.e.\ }}

\newcommand{\rmi}{\mathrm{i}}

\newcommand{\mC}{\mathcal{C}}
\newcommand{\mO}{\mathcal{O}}

\newcommand{\mH}{\mathcal{H}}

\newcommand{\dket}[1]{|#1 \rangle \! \rangle}
\newcommand{\dbra}[1]{\langle \! \langle #1 |}

\newcommand{\dbraket}[2]{\langle \! \langle #1|#2\rangle \! \rangle}
\newcommand{\Renyi}[2]{R^{(2)}\qty(#1,#2)}

\newcommand{\LandauO}[1]{O\qty(#1)}

\newcommand{\opnorm}[1]{\lVert  #1  \rVert_{\rm{op}}}

\newcommand{\bTr}[1]{\mathrm{Tr}\qty[#1]}

\newcommand{\kCS}{\mathsf{k}}

\renewcommand{\section}[1]{\medskip \noindent \textit{#1}:---}

\begin{document}

\title{Strong-to-Weak Spontaneous Symmetry Breaking from Wormholes in Holography}

\author{Taishi Kawamoto}
\email{taishi.kawamoto@yukawa.kyoto-u.ac.jp}
\affiliation{Department of Physics, Kyoto University, Kyoto 606-8502, Japan}
\affiliation{The Hakubi Center for Advanced Research, Kyoto University,
Yoshida Ushinomiyacho, Sakyo-ku, Kyoto 606-8501, Japan
}

\author{Kenya Tasuki}
\email{kenya.tasuki@yukawa.kyoto-u.ac.jp}
\affiliation{Center for Gravitational Physics and Quantum Information,\\
Yukawa Institute for Theoretical Physics, Kyoto University, 
Kyoto 606-8502, Japan
} 
 
\author{Masahito Yamazaki}
\email{masahito.yamazaki@ipmu.jp}
\affiliation{
    Graduate School of Physics,
    University of Tokyo, Tokyo 113-0033, Japan
    }
\affiliation{
    Kavli Institute for the Physics and Mathematics of the Universe,\\
    UTIAS, University of Tokyo, Kashiwa, Chiba 277-8583, Japan
    }
\affiliation{
    Trans-Scale Quantum Science Institute, 
    University of Tokyo, Tokyo 113-0033, Japan
    }
\affiliation{
    Center for Interdisciplinary Theoretical and Mathematical Sciences, RIKEN, Saitama 351-0198, Japan
    }

\begin{abstract}
We study strong-to-weak spontaneous symmetry breaking (SWSSB) at finite temperature and its implications in holography. SWSSB is a novel critical phenomenon that arises only in mixed states. We demonstrate SWSSB with the BTZ black hole and show that, in the bulk, it is realized through wormhole geometries connecting the two copies in the thermofield-double construction. We point out that SWSSB can be recast as an emergence of global symmetries under ensemble averaging. We further examine the interplay between SWSSB and the swampland program, and propose a new swampland conjecture for quantum-gravity theories involving ensemble averages.
\end{abstract}

\preprint{KUNS-3113, YITP-26-87, RIKEN-iTHEMS-Report-26}

\maketitle

\section{Introduction}
Symmetry and its spontaneous breaking are cornerstones of physics.
How a symmetry is realized, however, becomes markedly richer once a system is described by a density matrix rather than by a pure state, as happens, for example, in open systems and thermal ensembles.

A symmetry for a pure state $|\psi\rangle$ is described by a unitary operator $\hat{U}$ with action
$\hat U|\psi\rangle\propto|\psi\rangle$, and this preserves the density
matrix $\hat{\rho}$ as $\hat U\hat\rho\hat U^{\dagger}=\hat\rho$. This is a
\emph{weak symmetry}. A mixed state admits a sharper notion,
$\hat U\hat\rho=e^{\rmi\theta}\hat\rho$, in which $\hat U$ acts on
$\hat\rho$ as a mere phase; this is a \emph{strong symmetry}. A strong
symmetry is automatically weak, but not conversely---a distinction that
becomes physical precisely when the system interacts with its
environment \cite{Buca:2012zz,Albert:2014yej,Sieberer2016a}.

The distinction between strong and weak symmetries allows a symmetry-breaking pattern with no closed-system
analog: \emph{strong-to-weak spontaneous symmetry breaking} (SWSSB), in
which the choice of density matrix breaks a strong symmetry down to a
weak one. 
Implicitly in the literature for a long time, SWSSB has become
an active subject only recently~\cite{Minami:2015uzo,Hongo:2018ant,Hidaka:2019irz,Lee:2023fsk,Ma:2023rji,Lessa:2024gcw,Wang:2026ndj}. Related directions include quantum chaos and approximate
SWSSB~\cite{Saad:2018bqo,Khramtsov:2020bvs, Winer:2021nvd}, effective field theory of 
hydrodynamics~\cite{Crossley:2015evo,Ogunnaike:2023qyh,Akyuz:2023lsm,Huang:2024rml} and its relation to holography~\cite{deBoer:2018qqm,Glorioso:2018mmw}, and SymTFT~\cite{Schafer-Nameki:2025fiy}. 

In this Letter, we show that SWSSB is a unifying phenomenon across widely disparate setups, 
arising both in finite-dimensional lattice systems and in conformal field theories with holographic gravity duals---the latter provide canonical examples of strongly-coupled quantum field theories \cite{Maldacena:1997re,Aharony:1999ti}. In the gravitational setting, it acquires a geometric meaning: a bulk
realization of SWSSB in the spirit of the ER=EPR correspondence~\cite{Maldacena:2013xja}. Indeed, we will show that the long-range order in the Schwinger--Keldysh (SK) path integral, which characterizes the SWSSB, leads to geometric connections in the bulk gravitational path integrals. This implies that the SWSSB diagnoses the existence of a wormhole, which is a fundamental correlation and entanglement pattern of the gravitational path integral.

\section{SWSSB via R\'{e}nyi-$2$ and Wightman Correlators}
In quantum information theory and non-equilibrium physics, it is useful to regard a density matrix as a vectorized Choi state in the doubled Hilbert space $\mH_+\otimes\mH_-$, 
\begin{equation}
\label{eq:rho_doubled}
    \dket{{\rho}}:= \sum_{i}p_i\ket{\psi_i}_+\otimes\ket{\psi_i^*}_-,
\end{equation}
where each  $\mH_{\pm}$ is a copy of the original Hilbert space $\mH$.

Let us consider a charged local operator $\hat{\mO}(x)$ with charge $q\ne 0$ and the following state perturbed by the operator and its conjugate: 
\begin{equation}
 \dket{\sigma}:= {\hat{\mO}}_+(x){\hat{\mO}}_+(y)^\dag {\hat{\mO}}_- (y) {\hat{\mO}}_- (x)^\dag \dket{\rho}.
\end{equation}
The R\'{e}nyi-2 correlator is defined as the overlap between these two states \cite{Lee:2023fsk,Lessa:2024gcw} \footnote{%
In the literature, there are two distinct measures for SWSSB:
R\'{e}nyi-2 correlators and fidelity correlators.
It is known that the fidelity correlator $F(x, y)$ is bounded above
and below by the Wightman correlator
\protect\cite{Liu:2024mme,Weinstein:2024fug,Wang:2026ndj},
\begin{equation*}
    C^W(x,y) \leq F(x,y)\opnorm{\mO}^2,
    \quad
    F(x,y)\leq \sqrt{C^W(x,y)}. 
\end{equation*}
See also
\protect\cite{Liu:2026lhj,Divi:2026zrd,Zhang:2026nrp}
for recent papers on local characterization of SWSSB.
While the operator norm is UV divergent in quantum field theories,
we can resolve this divergence by introducing smearing in the spatial direction,
\ie
$\mO(x)\mapsto \mO[f]:=\int \dd^d x\, f(x) \mO(x)$.
Note that operators in bosonic lattice systems and in quantum field theories
suffer from divergences arising from the infinite dimensionality of local Hilbert
spaces or the noncompactness of the target space of fields.%
}%
\nocite{Liu:2024mme,Weinstein:2024fug,Wang:2026ndj,Liu:2026lhj,Divi:2026zrd,Zhang:2026nrp}
\begin{equation}\label{eq:Renyi-2}
 \begin{split}
     \Renyi{x}{y}&:=\frac{\dbraket{\rho}{\sigma}}{\dbraket{\rho}{\rho}}
     =\frac{\bTr{{\hat{\mO}}(x){\hat{\mO}}(y)^\dag {\hat{\rho}} {\hat{\mO}}(y){\hat{\mO}}(x)^\dag{\hat{\rho}}}}{\bTr{{\hat{\rho}}^2}}.
 \end{split}
\end{equation}
We also define the Wightman correlator (also known as the R\'{e}nyi-1 correlator) by
  \begin{equation}
      C^W(x,y) := \bTr{\hat{\mO}(x)\hat{\mO}(y)^\dag \sqrt{\hat{\rho}}\hat{\mO}(y)\hat{\mO}(x)^\dag\sqrt{\hat{\rho}}}.
  \end{equation}
  
Suppose we have a QFT with a symmetry group $G$, and a mixed state described by a density matrix ${\hat{\rho}}$ with a strong $G$ symmetry.
For concreteness, we restrict to the case $G=\mathrm{U}(1)$ in the following.
The mixed state ${\hat{\rho}}$ has SWSSB if there exists a charged local operator ${\hat{\mO}}(x)$ such that (a) its R\'{e}nyi-2 correlator is non-vanishing \footnote{%
One should notice that there are in general differences between SSB and
Long-Range-Order (LRO) \protect\cite{Tasaki_2018}, and the definition above is
the Strong-to-Weak analogue of the LRO, and hence should rather be called
``Strong-to-Weak LRO'' or ``R\'{e}nyi-2 LRO.''
We will, however, use the terminology SWSSB to simplify our discussion and to
match the literature.%
}%
\nocite{Tasaki_2018}
\begin{equation}
    \label{eq:SWSSB}
    \lim_{\abs{x-y}\to\infty} R^{(2)}(x,y)   = \LandauO{1},
\end{equation}
or if a similar equation holds for the Wightman correlator $C^W(x,y)$,  
while (b) the conventional correlation function shows no long-range order for any charged operators:
\begin{equation}
    \lim_{\abs{x-y}\to\infty}\bTr{ {\hat{\rho}} \, {\hat{\mO}}(x){\hat{\mO}}(y)^\dag}=0,
\end{equation}
so that the weak symmetry is not spontaneously broken.

\section{SWSSB at finite temperature}
In the following, we focus on SWSSB at finite temperature.
Following Ref.~\cite{Lessa:2024gcw}, we consider the following mixed state:
\begin{equation}
\label{eq:gmc}
    {\hat{\rho}}_{Q, \beta} = 
    {\hat{\Pi}_Q e^{-\beta \hat{H}}}/{\bTr{ \hat{\Pi}_Q e^{-\beta \hat{H}}}}, 
 \end{equation}
where $\hat{H}$ is the Hamiltonian of the system, $\beta\in(0,\infty)$ is the inverse temperature, and $\hat{\Pi}_Q$ is the projection operator to the sector with fixed charge $\hat{Q}=Q \in \mathbb{Z}$. This state is strongly symmetric. By contrast, the grand canonical ensemble 
\begin{equation}
    {\hat{\rho}}_{\mu,\beta} = {e^{-\beta({\hat{{H}}}-\mu \hat{Q})}}/{\bTr{e^{-\beta(\hat{H}-\mu \hat{Q})}}}
\end{equation}
has only a weak symmetry. 
Since the projection operator $\hat{\Pi}_Q$ commutes with the charge-neutral operator ${\hat{\mO}}(x){\hat{\mO}}(y)^{\dag}$, we can rewrite the R\'{e}nyi-$2$ correlator and Wightman correlator as 
\begin{equation}
\begin{split}
    R^{(2)}(x,y)&= 
\ev{{\hat{\mO}}(x;-\rmi\beta){\hat{\mO}}(y;-\rmi\beta)^{\dag}{\hat{\mO}}(y){\hat{\mO}}(x)^\dag}
_{Q, 2\beta},\\
C^W(x,y)&= \ev{{\hat{\mO}}\qty(x;-\rmi\frac{\beta}{2}){\hat{\mO}}\qty(y;-\rmi\frac{\beta}{2})^{\dag}{\hat{\mO}}(y){\hat{\mO}}(x)^\dag}
_{Q, \beta},
\end{split}
\end{equation}
where we denoted the expectation value as $\ev{\cdots}_{\alpha} := {\bTr{\hat{\rho}_\alpha \cdots}}/{\bTr{\hat{\rho}_\alpha}}$, and we obtained $\hat{\rho}_{Q, 2\beta}$
from $(\hat{\rho}_{Q, \beta})^2$.
With this choice of strongly symmetric state ${\hat{\rho}}$, the SWSSB is expected to hold in general local quantum many-body systems \cite{Lessa:2024gcw}:

\medskip\noindent
{\bf Lattice SWSSB Conjecture: }
Suppose we have a local $\mathrm{U}(1)$-symmetric Hamiltonian $\hat{H}$, temperature $0<\beta<\infty$, and a density matrix with a strong $\mathrm{U}(1)$ symmetry ${\hat{\rho}}_{Q, \beta}$.
If the density matrix ${\hat{\rho}}_{Q, \beta}$ \eqref{eq:gmc} has no SSB, then it must have SWSSB, i.e., \cref{eq:SWSSB} holds.

\medskip

While we will defer more rigorous proof of this conjecture for future work, let us recapitulate the reasoning behind it. We focus on the Rényi-2 correlator: a completely parallel discussion works for the Wightman correlator.

Since the weak symmetry is unbroken, we have 
\begin{equation}
\label{eq:one_two_vanish}
    \ev{{\hat{\mO}}(x)}_{Q, 2\beta} 
    =0,\; \lim_{\abs{x-y}\to\infty}\ev{{\hat{\mO}}(x){\hat{\mO}}^\dag(y)}_{Q, 2\beta} 
    =0.
\end{equation}
We can then invoke the cluster decomposition property for four-point functions
\footnote{For the present discussion, exponential decay of the correction terms is not required.}
to obtain
\begin{equation}
\label{eq:clustering}
    \begin{split}
        &\ev{{\hat{\mO}}(x;-\rmi\beta){\hat{\mO}}(y;-\rmi \beta)^{\dag}{\hat{\mO}}(y){\hat{\mO}}(x)^\dag}_{Q, 2\beta}
        \\
        & \approx \ev{{\hat{\mO}}(x;-\rmi\beta){\hat{\mO}}(x)^\dag}_{Q, 2\beta}
        \ev{{\hat{\mO}}(y;-\rmi \beta)^{\dag}{\hat{\mO}}(y)}
        _{Q, 2\beta}
    \end{split}
\end{equation}
in the limit $\abs{x-y}\to \infty$.
We further invoke the local ensemble equivalence for the two-point functions
\begin{equation}
    \label{eq:ensemble_equiv}
    \ev{{\hat{\mO}}(x;-\rmi\beta){\hat{\mO}}(x)^{\dagger}}_{Q, 2\beta}
    \approx \ev{{\hat{\mO}}(x;-\rmi\beta){\hat{\mO}}(x)^{\dagger}}_{\mu, 2\beta},
\end{equation}
where the chemical potential $\mu$ is fixed by $Q$ and $\beta$ by the standard thermodynamic law. (Similar equivalence between usual canonical and microcanonical ensemble was proven for local translation-invariant Hamiltonians under moderate assumptions on thermodynamic functions and clustering property of correlation functions \cite{Brandao:2015hwy,Tasaki_2018:OLE2018,Kuwahara:2020chr}. It is a non-trivial question whether the assumptions are satisfied for a given system \cite{Kawamoto202x}.) Finally, assuming the reflection positivity:
\begin{equation}
    \label{eq:gcan_positive}
       \ev{
       {\hat{\mO}}(x;-\rmi\beta)
       {\hat{\mO}}(x)^{\dagger}
       }_{\mu, 2\beta} 
        \!\!=  \ev{{\hat{\mO}}\left(x;-\rmi\frac{\beta}{2}\right){\hat{\mO}}\left(x;\rmi\frac{\beta}{2}\right)^{\dagger}}
        _{\mu, 2\beta} 
        \!\!\!\!\!\!\!\! > 0.
\end{equation}
By combining these with Eq.~\eqref{eq:clustering}, we obtain SWSSB for the R\'{e}nyi-2 correlator.
Since the argument is general, \ie we can apply the results to a wider class of local quantum many-body systems possibly involving quantum field theories, one can ask if the same statement applies to the holographic theory, to which we turn next.

\section{Holographic SWSSB}
To illustrate our points in a simple setup, we consider the bottom-up AdS$_3$/CFT$_2$ correspondence
(see e.g.\ \cite{Kraus:2006wn} for a review).
We consider a $G=\mathrm{U}(1)$ symmetry, which is a global symmetry on the boundary holographic QFTs and a gauge symmetry in the bulk theory.  We take the bulk theory to be Einstein gravity and $\mathrm{U}(1)$ Chern-Simons theory with level $\kCS>0$. We also consider a complex bulk scalar field coupled to the $\mathrm{U}(1)$ Chern-Simons theory, which plays the role of a probe field to compute the Rényi-2 correlator.  Since the Chern-Simons theory is topological, we can solve the equations of motion for the metric and gauge fields independently.
We expect the Hawking-Page phase transition as we change the temperature \cite{Hawking:1982dh,Witten:1998zw}, and 
depending on the value of $\beta$, the relevant bulk metric in the large $N$ limit 
is either the Bañados–Teitelboim–Zanelli (BTZ) black hole \cite{Banados:1992wn} or thermal AdS. To discuss long-range order, we consider the infinite-volume limit. In this case, since the modular parameter is given by the ratio of inverse temperature $\beta$ and spatial volume $L$, \ie $\tau = \rmi {\beta}/{L}$, we always have a high-temperature regime, \ie the dominant contribution for the gravitational path integral is given by the BTZ black hole. The imaginary time evolution of the BTZ black hole is represented as the product of the interval  $[0, \beta]$ times the two-dimensional torus $T^2$.

We first invoke the large-$N$ factorization in our ensemble, which gives
\begin{equation}
\begin{split}
 \Renyi{x}{y}&\approx 
 \ev{{\hat{\mO}}(x;-\rmi\beta){\hat{\mO}}(x)^\dag}_{Q, 2\beta}
 \ev{{\hat{\mO}}(y;-\rmi \beta)^\dag{\hat{\mO}}(y)}_{Q, 2\beta}
 \\
 &
  + \ev{{\hat{\mO}}(x;-\rmi\beta){\hat{\mO}}(y;-\rmi\beta)^\dag}_{Q, 2\beta}
  \ev{{\hat{\mO}}(y){\hat{\mO}}(x)^\dag}_{Q, 2\beta}
\end{split}
\end{equation}
in the leading large $N$ limit.
This is a holographic counterpart of the cluster decomposition \eqref{eq:clustering} for lattice systems.
In the large distance limit $\abs{x-y}\to \infty$, the second term vanishes; hence, the only remaining task is to show that the holographic two-point function
$\ev{{\hat{\mO}}(x;-\rmi\beta){\hat{\mO}}(x)^\dagger}_{Q, 2\beta}
$ is non-vanishing. 
This is essentially a standard computation in holography, 
except that the precise bulk realization of the projection operator $\hat{\Pi}_Q$ is subtle. In what follows, we compute the Rényi-$2$ correlator using a holographic implementation of the charge projection.

\begin{figure}
    {\centering
    \begin{tikzpicture}[line width=0.7pt]

\draw[thick] (0,0) ellipse (0.7 and 1.3);

\draw[thick] (6,1.3) arc[start angle=90, end angle=-90, x radius=0.7, y radius=1.3];
\draw[thick, dashed] (6,-1.3) arc[start angle=270, end angle=90, x radius=0.7, y radius=1.3];

\draw[thick] (0,1.3) .. controls (3,0.6) .. (6,1.3);
\draw[thick] (0,-1.3) .. controls (3,-0.6) .. (6,-1.3);

\fill[green!70!black] (0.65,0.6) circle (2pt) ;
\draw (0.65,0.8)node[right]{\Large $y$};
\fill[green!70!black] (0.65,-0.6) circle (2pt);
\draw (0.65,-0.4)node[right]{\Large $x$};
\node[green!70!black] at (0.15,0.6) {\Large $\mathcal{O}^\dagger$};
\node[green!70!black] at (0.15,-0.6) {\Large $\mathcal{O}$};

\fill[green!70!black] (6.6,0.6) circle (2pt)  ;
\draw(6.6,0.8)node[left]{\Large $y$};
\fill[green!70!black] (6.6,-0.6) circle (2pt);
\draw(6.6,-0.4)node[left]{\Large $x$};
\node[green!70!black] at (7,0.6) {\Large $\mathcal{O}$};
\node[green!70!black] at (7,-0.6) {\Large $\mathcal{O}^\dagger$};

\draw[red,ultra thick] (3,0.8) .. controls (3.2,0) .. (3,-0.8);
\draw[red,ultra thick, dashed] (3,0.8) .. controls (2.8,0) .. (3,-0.8);

\node[red] at (3.5,0) {\Large $\mathcal{C}$};

\draw[green!70!black, thick] (0.65,-0.6) .. controls (3,-0.3) .. (6.6,-0.6); 
\draw[green!70!black, thick] (0.65,0.6) .. controls (3,0.3) .. (6.6,0.6);   
\end{tikzpicture}}
\caption{We represent four boundary local operators $\mathcal{O}, \mathcal{O}^{\dagger}$ and a bulk Wilson line wrapping a non-contractible cycle $\mathcal{C}$ representing the chemical potentials on the boundary. The horizontal direction represents Euclidean time evolution from $t_E=0$ to $t_E=\beta$.}
\label{fig:Wilson_line}
 \vspace{-7pt}
\end{figure}
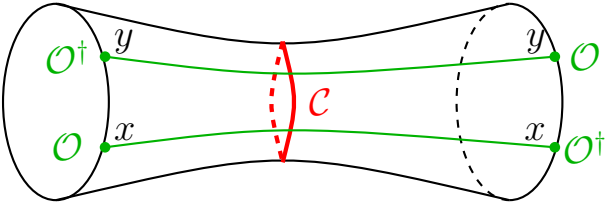

In CFTs with a $\mathrm{U}(1)$ global symmetry, we can express the projection operator by the following integral
over the (imaginary) chemical potential $\mu$:
\begin{equation}\label{projector}
    \hat{\Pi}_Q = \int_{0}^{2\pi} \frac{\dd \mu}{2\pi}e^{\rmi \mu(\hat{Q}-Q)}.
\end{equation}
In the gravity dual 
each operator 
$e^{\rmi \mu \hat{Q}}$ 
is identified as the Wilson line defect 
with non-trivial holonomy $\mu$
along a non-contractible cycle $\mathcal{C}$
\footnote{%
Ref.\ \protect\cite{Zhao:2020qmn} considered the holographic dual of a charge
cumulant to compute the symmetry-resolved entanglement entropy
\protect\cite{Goldstein:2017bua} for a subsystem $\mathcal{A}$,
and identified its bulk dual to be the Wilson line defect.
There, the bulk line $\mathcal{C}$ connects the entanglement surface
$\partial \mathcal{A}$ and is determined as the Ryu-Takayanagi surface
\protect\cite{Ryu:2006bv}.
For our purposes, we consider a finite-temperature system and replace the
subsystem $\mathcal{A}$ by the whole system.%
}%
\nocite{Zhao:2020qmn,Goldstein:2017bua,Ryu:2006bv}.
In the BTZ geometry, we can take $\mathcal{C}$ to be the non-contractible cycle wrapping the horizon of the BTZ black hole. See FIG.~\ref{fig:Wilson_line} for illustration.
We can then integrate over $\mu$ to implement the projection operator \eqref{projector}.

The Wilson line generates the modified boundary conditions for the complex scalar field:
\begin{equation}
    \Phi(t_E+\beta) = e^{-\rmi \mu q } \Phi(t_E),
\end{equation}
where $q$ is the $\mathrm{U}(1)$ charge for the probe field. This twisted boundary condition leads to a shift of the Matsubara frequency,
\begin{equation}
    \omega_n = \frac{2\pi}{\beta}\qty(n+\frac{\mu q}{2\pi}),\; n\in \mathbb{Z}.
\end{equation}
Following the standard dictionary of the AdS/CFT correspondence, we can solve the wave equation and evaluate the two-point functions under the twisted boundary conditions. The procedure is almost the same as computing the quasi-normal modes in the BTZ background \cite{Birmingham:2001pj,Birmingham:2001hc}. 
We find that the ordinary two-point function shows short-range correlation in the spatial direction and no conventional SSB. Moreover, by using the large $N$ factorization, we can show that the Wightman correlator and the Rényi-2 correlator are non-zero in the large distance limit,
\begin{equation}
   \lim_{\abs{x-y}\to \infty} \! C^W(x,y)\geq  \lim_{\abs{x-y}\to \infty}\! R^{(2)}(x,y) \geq \frac{2}{\pi}\qty(\frac{4\pi}{\beta})^4  \frac{359}{945},
\end{equation}
which demonstrates SWSSB in our holographic models.
We provide details in the Supplemental Material (see in particular \cref{eq:result} and \cref{eq:result2}).
We can repeat a similar analysis for the thermal AdS$_3$ geometry. 
In this case, the chemical potential should cause the holonomy of the gauge field along the 
temporal non-contractible cycle. Note that an exchange of spatial and temporal directions from the BTZ case is expected naturally from the modular transformation.  

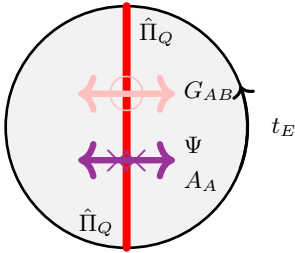
\begin{figure}[htbp]
\vspace{-7pt}
  \centering
  \begin{tikzpicture}[line cap=round,line join=round, scale=0.8]
    \def\R{2.0}

    \fill[lightgray!20] (0,0) circle (\R);

    \draw[line width=1.0pt] (0,0) circle (\R);

    \draw[red, line width=3.0pt] (0,\R) -- (0,-\R);

    \node[anchor=south] at (0.5,\R-0.8) {\small$\hat{\Pi}_Q$};
    \node[anchor=north] at (-0.5,+0.8-\R) {\small$\hat{\Pi}_Q$};

    \draw[->, line width=1.2pt]
      ({\R*cos(-20)},{\R*sin(-20)})
        arc[start angle=-20, end angle=20, radius=\R];

    \node[anchor=west]
      at ({(\R+0.25)*cos(0)},{(\R+0.25)*sin(0)}) {\small$t_E$};

    \node[anchor=west] at (0.8, 0.6) {\small$G_{AB}$};
    \node[anchor=west] at (0.8,-0.3) {\small$\Psi$};
    \node[anchor=west] at (0.8,-0.9) {\small$A_A$};

    \def\yA{0.55}
    \def\yB{-0.55}

    \draw[<->, line width=2.4pt, pink] (-0.8,\yA) -- (0.8,\yA);
    \node[pink] at (0,\yA) {\Large\bfseries$\bigcirc$};

    \draw[<->, line width=2.4pt, violet!80] (-0.8,\yB) -- (0.8,\yB);
    \node[violet!80] at (0,\yB) {%
      \Large\bfseries
      \raisebox{0pt}{\scalebox{1.2}{$\times$}}%
      \kern-0.45em%
      \raisebox{0pt}{\scalebox{1.2}{$\times$}}%
    };
  \end{tikzpicture}
  \vspace{-7pt}
  \caption{Strongly-symmetric configuration in the gravitational path integral dual to $\bTr{e^{-\beta \hat{H}}\hat{\Pi}_Q e^{-\beta \hat{H}}\hat{\Pi}_Q}$.
  Neutral fields such as the graviton can pass the brane (red) freely,
  while the charged fields $\Psi$ and gauge fields $A$ (purple) cannot pass.}
  \label{fig:Pi_Q}
\end{figure}

\begin{figure}
\vspace{-12pt}\centering\includegraphics[width=\linewidth]{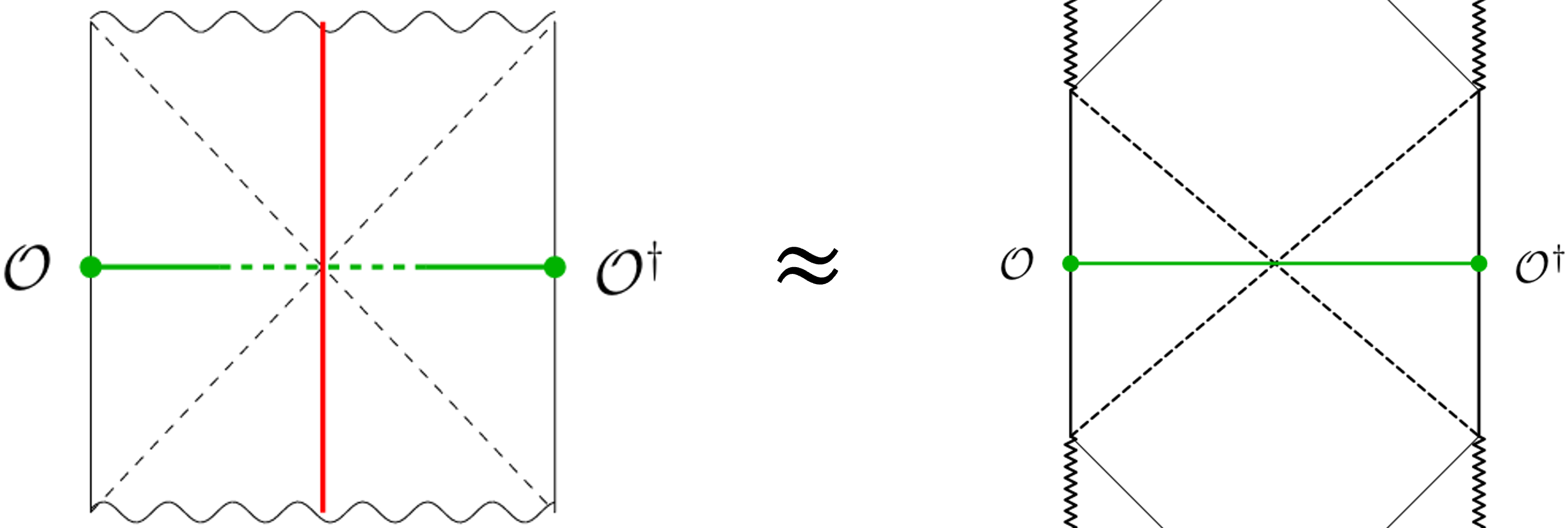}
 \vspace{-12pt}
    \caption{The bulk geometry with a defect will be approximated by the charged black hole solution. The latter does not have any obstacles for charged fields to communicate.}
    \label{fig:Approximations}
\end{figure}

The essence behind the non-vanishing of Rényi-2 and Wightman correlators is the spacetime connectivity and the non-zero contribution of the factorized channel of the correlators. We stress, however, that such a connectivity is not obvious at first sight.
To further discuss this point, let us examine the property of the projection operator $\hat{\Pi}_Q$ and the gravity configuration that possesses a strong symmetry. On the boundary side, the projection operator plays the role of a co-dimension one defect such that a charged operator cannot pass through but a neutral operator can, as guaranteed by the formula
\begin{equation}
    \hat{\Pi}_{Q'} \hat{\mO}_q \hat{\Pi}_Q = \delta_{Q',Q+q} \hat{\mO}_q \hat{\Pi}_Q.
\end{equation}
From the AdS/BCFT (Boundary CFT) \cite{Takayanagi:2011zk,Fujita:2011fp} and AdS/ICFT (Interface CFT) correspondence \cite{Bak:2003jk,Clark:2004sb,Bak:2007jm,Gutperle:2018fkz}, the gravity dual of a strongly symmetric theory is naturally modeled by a bulk codimension-one brane opaque to the bulk complex charged fields and gauge fields. In this picture, we may imagine that the one-point function and left-right two-point function are zero, which leads to vanishing Rényi-2 correlators, see FIG.~\ref{fig:Pi_Q}. This is indeed the manifestation that the state $\hat{\rho}_{Q,\beta}$ is strongly symmetric: we can freely perform the gauge transformation on the left and right of the brane independently. Since we know the left-right two-point functions and Rényi-2 correlators are non-zero, we should have a bulk geometry that connects the left and right such that the charged fields can communicate in some manner. One possible resolution is that we have a wormhole geometry in the bulk. From the discussion in the Supplemental Material, we physically expect that our strongly symmetric state is approximated by the grand canonical ensemble as a consequence of the ensemble equivalence. The gravity dual of the grand canonical ensemble will be a charged black hole (see FIG.~\ref{fig:Approximations}). These charged black holes do not pose any obstacle for charged fields, so that the two asymptotic boundaries are connected. In other words, the weak symmetry can be related to the appearance of wormholes.

In the holographic computations above, we expect the correlators to be non-vanishing even when we have a finite number of degrees of freedom (large but finite $N$ or $c$ in holographic theories) and a finite volume since the spacetime is still connected in this regime.  While we should not expect strict SWSSB in finite degrees of freedom, we see that other saddle points such as  $\mathrm{SL}(2,\mathbb{Z})$ black hole solutions and thermal AdS solutions still connect two SK boundaries (at least as Euclidean geometries) and thus break strong symmetry in each configuration. To obtain unbroken $G\times G$ symmetry, for example, we need to find brane-type solutions that separate the charged sectors of the two boundaries discussed above in finite $c$ and finite volume. We may say this is a ``symmetry-resolved version'' of the information loss paradox \cite{Maldacena:2001kr,Barbon:2003aq,Kleban:2004rx} in the sense that we lose information about the superselection sector of the full $G\times G$ symmetry \footnote{This discussion is naive in that we implicitly made a non-trivial identification between long-range order and SSB. The information-loss problem discussed above pertains to SWSSB rather than long-range order.}.

\section{Weak Symmetry as Emergent Ensemble Symmetries}
We next turn to conceptual points concerning SWSSB and wormholes from the point of view of ensemble averaging. In the literature, wormholes in gravitational path integrals are often related to ensemble averages over some randomness \cite{COLEMAN1988867,GIDDINGS1988890,VanRaamsdonk:2010pw,Cotler:2016fpe,Saad:2019lba,Kundu:2021nwp,Belin:2020jxr,Kawamoto:2024vzd}. Thus, it is natural to argue that a weak symmetry in SWSSB can be recast as an emergent symmetry in ensemble averages, discussed recently, e.g., in Refs.~\cite{Ashwinkumar:2023jtz,Antinucci:2023uzq} \footnote{The relation of SWSSB to 
spin-glass order was discussed briefly in Ref.~\cite{Lessa:2024gcw}.}.
This may be a perplexing statement since SWSSB is a symmetry breaking from a more symmetric to a less symmetric state, while, by contrast, the emergent symmetry in an ensemble average is an opposite phenomenon wherein we obtain a more symmetric state from a less symmetric state. We will, however, argue below that these two perspectives are complementary and describe the same physical phenomena.

We consider a weak-symmetric state obtained by the group average (also known as twirling):
 \begin{equation}
     \rho = \frac{1}{\abs{G}}\int \dd \mu(g) \,
     \hat{U}_g\ket{\psi}\bra{\psi}\hat{U}_g^\dag,
 \end{equation}
where $\dd\mu(g)$ is the Haar measure, and we can consider a generic non-symmetric state $\ket{\psi}$. 
Then we consider the SK path integral
\begin{equation}
    \begin{split}
        &\quad \bTr{e^{-\rmi \hat{H}t}\hat{\rho} \; e^{\rmi \hat{H}t}}\\
        &= \frac{1}{\abs{G}}\int \dd \mu(g) \sum_{\alpha} \bra{\alpha}e^{-\rmi \hat{H}t}\hat{U}_g\ket{\psi} \bra{\psi}U_g^\dag e^{\rmi \hat{H}t}\ket{\alpha}\\
        &=  \frac{1}{\abs{G}}\int \dd \mu(g) \sum_{\alpha} Z_g(\alpha,\psi) Z_g(\alpha,\psi)^*,
    \end{split}
\end{equation}
 where $\ket{\alpha}$ is an orthogonal basis and we define 
 \begin{equation}
     Z_g(\alpha,\psi) := \bra{\alpha}e^{-\rmi \hat{H}t}\hat{U}_g\ket{\psi}.
 \end{equation}
 Note that each path integral $Z_{g}(\alpha,\psi)$ does not possess the symmetry since the boundary condition related to $\ket{\psi}$ explicitly breaks the symmetry. The original weak symmetry can now be regarded as an emergent symmetry.
 
This emergence can also be explained in the SK contour in the path integral formalism (cf.\ \cite{Haehl:2017qfl,Haehl:2017eob}):
\begin{equation}
\label{eq:SK}
\begin{split}
    e^{W[J_+,J_-]} &=
\int_{\rho} D\phi_+D\phi_- e^{\rmi S_+[\phi_+,J_+]-\rmi S_{-}[\phi_-,J_-]},
\end{split}
\end{equation} 
where the information of the initial state is encoded in the junction conditions which interpolate between the double copies $\phi_+(0)$ and $\phi_-(0)$. The SK action $\rmi S_+[\phi_+,J_+]-\rmi S_{-}[\phi_-,J_-]$ enjoys a doubled symmetry $G_+\times G_-, \;G_{\pm}\cong G$.
This is broken by the choice of mixed state $\rho$ relating $\phi_+$ and $\phi_-$.

For concreteness, let us consider a mixed state perturbed by an operator deformation which keeps weak symmetry but not strong symmetry,
\begin{equation}
   \dket{\rho}=\exp\qty(\sum_{q}\lambda_q\int_{M} \, \mO_+^{[q]}{\mO}_-^{[-q]})\ket{\psi_+}\ket{\varphi_-},
\end{equation}
where $\ket{\psi_+}$ and $\ket{\varphi_-}$ are invariant under symmetry actions. The condition that $\dket{\rho}$ is weak-symmetric is reflected by the fact that this deformation keeps the diagonal symmetry, while each field $\mO_{\pm}^{[\pm q]}(x)$ breaks the symmetry. The total path integral is given by
\begin{equation}
\begin{split}
    e^{W[J_+,J_-]}&=\int D\phi_{\pm} \, e^{\rmi  S[\phi;J]}\\
    S[\phi;J] &= S_+[\phi_+;J_+]-S_-[\phi_-;J_-]\\
    &- \rmi \sum_{q}\lambda_q\int_{M}
    \dd^d x
    \, \mO_+^{[q]} {{\mO}}_-^{[-q]},
\end{split}
\end{equation}
which can be rewritten by introducing the Hubbard–Stratonovich (HS) transformation as
\begin{equation}
    \begin{split}
        e^{W[J_+,J_-]} &= \int D\alpha_{+} D\alpha_{-}\;P(\alpha_{+},\alpha_{-})\;
        \prod_{\sigma=\pm} Z_{\sigma}[J_{\sigma};\alpha_{\sigma}], \\
        P(\alpha_{+}, \alpha_{-}
        )&\propto \exp\qty(\sum_q \lambda_q^{-1}\int \dd^d x 
        \, \alpha_{+}^{[q]}\alpha_{-}^{[q]}),\\
        Z_\pm[J_{\pm};\alpha_{\pm}]& \!:= \! \int \!\! D\phi_\pm \exp\!\qty(\!\pm \rmi(S_{\pm}\!+\!\sum_q \!\int\! \alpha_{\pm}^{[q]}\mO_{\pm}^{[\pm q]})\!).
    \end{split}
\end{equation} 
Each $Z_{\pm}[J_{\pm};\alpha_{\pm}]$ does not possess the symmetry $G$, which is explicitly broken by the deformations. We find, however, that we restore the diagonal symmetry by averaging over the probability distribution $P(\alpha_{+}, \alpha_{-})$. Consequently, the symmetry emerges through the ensemble averaging.
We can repeat a similar analysis for more generic weakly symmetric states and again reinterpret the result as emergent symmetries in ensemble averages. Notice that the discussion is parallel to the discussion that the effective field theory for wormholes either breaks symmetries explicitly or induces spontaneous symmetry breaking \cite{COLEMAN1988867,GIDDINGS1988890,Abbott:1989jw,Hsin:2020mfa}. We may understand SWSSB by the following mechanism: first, the strong symmetry is completely broken; a weak symmetry then emerges after ensemble averaging. From a holographic perspective, this randomness and the emergent weak symmetry will be related to the semi-classical wormhole.

\section{Interplay with the Swampland Program}
Since our discussion above is general, it is natural to imagine that our conclusion can be recast as a more general guiding principle on SWSSB in general theories of quantum gravity (QG), in the spirit of the swampland program \cite{Vafa:2005ui,Ooguri:2006in}.

One should note, first and foremost, that one of the most well-known swampland conjectures states that there are no exact global symmetries in theories of QG \cite{Banks:1988yz,Banks:2010zn,Harlow:2018tng,Fichet:2019ugl,Daus:2020vtf,King:2023ztb,Dvali:2018txx,Dvali:2017eba}, and even a weak symmetry is forbidden by this conjecture. 
While this conjecture is often believed to be true,
in the literature we also encounter ensemble-averaged theories in the discussions of holography, 
where we encounter emergent global symmetries after such averaging \cite{Ashwinkumar:2021kav,Ashwinkumar:2023jtz,Antinucci:2023uzq}. While the necessity of such ensemble averaging is under debate, they do seem to arise in some simple theories of gravity, e.g.\ in two spacetime dimensions \cite{Saad:2019lba,Stanford:2019vob}, 
and may be needed for the resolution of the factorization puzzle \cite{Witten:1999xp,Maldacena:2004rf}.

One may therefore be tempted to extend the swampland global symmetry conjecture to 
more general holographic theories involving ensemble averages. To fill in this gap, we propose the following conjecture by taking advantage of the connection between emergent symmetries and SWSSB mentioned above.

{\bf Swampland SWSSB Conjecture:}
Consider any theory of quantum gravity with a description as a weakly-coupled Einstein gravity, with or without ensemble averages in the description. Then any strong global symmetry of a mixed state should be broken at least down to a weak symmetry; namely, the state should exhibit either SWSSB or strong SSB. It would be interesting to either prove or find a counterexample to this conjecture in string theory compactifications.

Let us next discuss another swampland conjecture \cite{Aspinwall:1995zi,Brennan:2017rbf}: the number of massless fields in a $(d+1)$-dimensional effective field theory coupled to Einstein gravity must be bounded from above by a number $N_{\text{max}}$ that depends only on the dimension $d+1$. 

Consider a quantum state $\dket{\rho}$ that lives in an extended Hilbert space $\mH^{\otimes n}$, wherein we can consider a generalization of strong and weak symmetry for the group $G^n=G\times G \times\cdots \times G$. We consider a SK amplitude $\dbra{\rho}\exp(\rmi \sum_{i=1}^n H_i t)\dket{\rho}$, where each Hamiltonian $H_i$ possesses the $G$ symmetry. In the dictionary of the AdS/CFT correspondence, we have a bulk dual for each contour \cite{Skenderis:2008dh,Skenderis:2008dg} and correspondingly $n$ copies of vector massless gauge fields \cite{Ferrara:1998jm,Kaplan:AdSCFTBottomUp}. The conjecture now excludes $n > N_{\text{max}}$. To reduce the number of bulk massless gauge fields, we should have either wormholes of the Einstein-Rosen type or the bulk Higgsing \cite{Aharony:2006hz,Karch:2023wui}. 
In the former case, in the Euclidean signature, we have a single connected geometry and a single massless gauge field from the outset. After analytic continuation, we get two boundaries, but the number of gauge fields is essentially still one. The corresponding boundary symmetry is therefore only the diagonal, weak symmetry.  In the latter case, the boundary states should spontaneously break the larger strong symmetry $G^n$.

In this Letter, we discussed a fascinating interplay between SWSSB 
and spacetime connectivity and the emergence of symmetries in holography.
Our discussion further strengthens the deep conceptual connections between entanglement and geometry via the lens of symmetry. In the AdS/CFT correspondence, the global symmetries of boundary QFTs are realized as the asymptotic symmetries in the bulk. Thus, the Goldstone modes in the boundary should be related to these asymptotic symmetries.  
One possible realization is in terms of relative gauge fields associated with the asymptotic symmetries of the two boundaries, similar to those in Refs.~\cite{deBoer:2018qqm,Saad:2018bqo,Chen:2023hra}.
It is tantalizing that the SWSSB, originally motivated by non-equilibrium phenomena in open systems, has deep implications in quantum gravity.

\section{Acknowledgements}
We would like to thank Satoshi Iso, Jong Yeon Lee, Keito Shimizu, Tadashi Takayanagi, and Zixia Wei for stimulating discussions.
KT is supported by Grant-in-Aid for JSPS Fellows No.\ 25KJ1455.
MY is supported in part by the World Premier International Research Center Initiative (WPI), MEXT, Japan; by the JSPS KAKENHI Grant No.\ 23K25865; by JST, Japan (CREST Grant No.\ JPMJCR26XA, Moonshot R\&D Grant No.\ JPMJMS2061); and by the IBM-UTokyo-sponsored research.
\bibliographystyle{apsrev4-1}
\bibliography{SWSSB}

\clearpage
\appendix
\onecolumngrid
\widetext

\setcounter{equation}{0}
\setcounter{figure}{0}
\setcounter{table}{0}
\setcounter{page}{1}
\makeatletter
\renewcommand{\theequation}{S\arabic{equation}}
\renewcommand{\thefigure}{S\arabic{figure}}
\renewcommand{\bibnumfmt}[1]{[S#1]}

\begin{center}
\textbf{\large Supplemental Material}
\end{center}

In this supplementary material, we describe the details of holographic computation of the R\'{e}nyi-2 correlator and the Wightman correlator for a charged light operator $\mO$ in a two-dimensional holographic CFT.
The total charge operator $\hat{Q}$ decomposes into holomorphic and anti-holomorphic parts: $\hat{Q} = \hat{Q}_{L}+\hat{\bar{Q}}_{R}$.
We denote the total charge eigenvalue of the probe operator $\mO$ as $q=q_{L}+\bar{q}_{R}$.

A bottom-up model of a two-dimensional holographic CFT with a global $\mathrm{U}(1)$ symmetry is given by the following Einstein-Chern-Simons theory coupled with probe matter~\cite{Kraus:2006nb,Kraus:2006wn,Zhao:2020qmn},
\begin{equation}
\begin{split}
    I_{\mathrm{grav}}[G,A,\bar{A},\Psi] &= I_{\mathrm{EH+GH}}[G] + I_{\mathrm{CS}}[A,\bar{A}] + I_{\mathrm{scalar}}[\Psi;G,A,\bar{A}],
    \\
    I_{\mathrm{EH+GH}}[G] & = \frac{1}{16\pi G_N}\int_{\mathrm{bulk}}\dd^3 x \sqrt{G}\qty[R[G]+2\Lambda] + \frac{1}{8\pi G_N}\int_{\partial \mathrm{bulk}} \dd^2 x \sqrt{h}(K[h]-1),\\
    I_{\mathrm{CS}}[A,\bar{A}]&= \frac{\rmi \kCS}{8\pi}\int_{\mathrm{bulk}} \qty(A \wedge \mathrm{d}A-\bar{A} \wedge \mathrm{d}\bar{A})-\frac{\kCS}{16\pi}\int_{\partial\mathrm{bulk}} \sqrt{h}h_{ab}(A^aA^b +\bar{A}^a\bar{A}^b)-\frac{\rmi \kCS \mu}{2\pi}\int_\mC (A-\bar{A}), \\
    I_{\mathrm{scalar}}[\Psi;G, A,\bar{A}] &= \int_{\mathrm{bulk}} \dd  ^3 x \sqrt{G} \qty[-D_A \Psi (D^A \Psi)^\dag-M^2\abs{\Psi}^2],
\end{split}
\end{equation}
where $\kCS>0$ is the Chern-Simons level, $h$ is the induced metric on conformal boundary and $D_A :=\nabla_A-\rmi q_{L} A_A-\rmi \bar{q}_{R}\bar{A}_A$ is the covariant derivative.
The bulk complex scalar $\Psi$ plays the role of the probe field dual to $\mO$. Since the Chern-Simons action is topological, we can solve the bulk equation of motion for gravity independently.
Thus, we can consider the BTZ black hole, which is the dominant saddle point in the large-volume limit for any non-zero temperature.
We discuss the BTZ black hole with inverse temperature $\beta$ in the coordinate system
\begin{equation}
\dd s^2
= \frac{r_0^2 z}{1-z}\, \dd t_E^2
+ \frac{\dd z^2}{4z(1-z)^2}
+ \frac{r_0^2}{1-z}\, \dd  x^2 ,
\qquad
\beta=\frac{2\pi}{r_0},
\qquad
t_E \sim t_E + \beta,
\qquad
x \sim x+ L.
\end{equation}
Later we take $L\to \infty$. Similar to the holographic entanglement entropy formula and its analogue for a Wilson line, the line operator wraps the black hole horizon \cite{Ammon:2013hba}. Thus, we identify the non-contractible cycle $\mC$ for the holonomy of the gauge field as a bifurcation surface of the BTZ black hole. The equations of motion of Chern-Simons gauge fields are 
\begin{equation}
\dd A=2\mu \star\delta_{\mC},
\qquad
\dd\bar A=2\mu \star\delta_{\mC},
\end{equation}
where $\delta_\mathcal{C}$ is the current one-form localized on $\mathcal{C}$. This leads to the following holonomy \cite{Zhao:2020qmn},
\begin{equation}
    \int_{\partial \Sigma}A = \int_{\partial \Sigma}\bar{A} =2\mu ,
\end{equation}
where $\Sigma$ is an oriented codimension-one surface intersecting
$\mathcal C$ transversely once, or equivalently, whose boundary
$\partial\Sigma$ links $\mathcal C$ once. In the gauge-fixing condition $A_{z}=\bar{A}_z=0$, the gauge fields realizing this holonomy are given by 
\begin{equation}
A=\frac{2\mu}{\beta}\,\dd t_E,
\qquad
\bar A=\frac{2\mu}{\beta}\,\dd t_E .
\end{equation}
From these solutions, we can evaluate the partition function from the on-shell action,
\begin{equation}
    \bTr{e^{\rmi \mu\hat{Q}}
    e^{-\beta \hat{H}}} \sim \sum_{m\in\mathbb{Z}}\exp(\frac{\pi cL}{6\beta}-\frac{\kCS L (\mu+2m\pi)^2}{2\pi\beta}).
\end{equation}
We need to sum over the spectral flow $\mu\to \mu+2\pi$ to get correct periodicity in $\mu$ for the integer charge spectrum of $\hat{Q}$, which corresponds to the large gauge transformation of bulk gauge fields. 
The expression of the partition function is known in the black hole phase \cite{Kraus:2006nb,Kraus:2006wn}. The $\mu$ integral of this partition function gives the partition function of the pure strongly symmetric Gibbs state with charge projection 
\begin{equation}
    \bTr{\hat{\Pi}_Q e^{-\beta \hat{H}}} \sim \int_0^{2\pi}\frac{\dd \mu}{2\pi}\;\sum_{m\in \mathbb{Z}}\exp(\frac{\pi c L}{6\beta}-\frac{\kCS L (\mu+2\pi m)^2}{2\pi\beta}-\rmi \mu Q) = \sqrt{\frac{\beta}{2\kCS L}}\exp\qty(\frac{\pi c L}{6\beta}-\frac{\pi \beta Q^2}{2\kCS L}).
\end{equation}
 Next, we discuss the thermodynamic functions of the grand canonical ensemble,
 \begin{equation}
     \hat{\rho}_{\mu,\beta} = \frac{e^{-\beta(\hat{H}-\mu\hat{Q})}}{\bTr{e^{-\beta(\hat{H}-\mu\hat{Q})}}},\; \psi_L(\mu,\beta):= \frac{1}{L}\log{\bTr{e^{-\beta(\hat{H}-\mu\hat{Q})}}}.
 \end{equation}
 The grand canonical ensemble is obtained by the analytic continuation of $\mu$ in the ensemble $\bTr{e^{\rmi \mu \hat{Q}-\beta \hat{H}}}$, $\mu \mapsto -\rmi {\beta \mu}$. In the large $L$ limit, the $m$-sum is dominated by $m=0$ and we can analytically continue to 
 \begin{equation}
\psi_L(\mu, \beta) = \frac{\pi c}{6\beta} + \frac{\kCS \mu^2}{2\pi \beta}+ \LandauO{\frac{\log{L}}{L}}.
 \end{equation}
By using the naive saddle point analysis, we can confirm the ensemble equivalence
 \begin{equation}
 \begin{split}
   \bTr{e^{-\beta(\hat{H}-\mu\hat{Q})}} &= \sum_Q e^{\beta \mu Q}\bTr{\hat{\Pi}_Qe^{-\beta \hat{H}}}\\
     &\overset{\mathrm{q}:=Q/L}{\approx} \int \dd \mathrm{q} e^{\beta L\mu \mathrm{q}} e^{\frac{\pi c L}{6\beta}-\frac{\pi\beta \mathrm{q}^2 L}{2\kCS}}\approx \exp\qty(L \sup_{\mathrm{q}}\qty[\beta \mu \mathrm{q}+\frac{\pi c}{6\beta}-\frac{\pi \mathrm{q}^2}{2\kCS}]).
 \end{split}
 \end{equation}
 We can easily confirm
 \begin{equation}
     \psi(\mu,\beta) = \sup_{\mathrm{q}}\qty[\beta \mu \mathrm{q} +\frac{\pi c}{6\beta}-\frac{\pi \mathrm{q}^2}{2\kCS}] = \frac{\pi c}{6\beta}+\frac{\kCS\mu^2}{2\pi \beta},
 \end{equation}
 which is deduced from the concavity of the thermodynamic function $\frac{1}{L}\log{\bTr{\hat{\Pi}_Q e^{-\beta \hat{H}}}}=\frac{\pi c}{6\beta}-\frac{\pi q^2}{2\kCS}$. This discussion leads to the ensemble equivalence for the thermodynamic partition functions.
The ensemble equivalence for correlation functions needs further information on correlation functions \cite{Tasaki_2018:OLE2018} and we leave a more detailed discussion of this issue for future work~\cite{Kawamoto202x}.

Next, we solve the wave equation for the probe field,
\begin{equation}
(D_A D^A - M^2)\Psi = 0.
\end{equation}
By a gauge transformation analogous to the Aharonov--Bohm effect, it is convenient to rotate the frame,
\begin{equation}
\Psi(P)=\exp\!\left(\rmi q_{L} \int_{P_{\infty}}^{P} A+\rmi \bar{q}_{R} \int_{P_{\infty}}^{P} \bar{A}\right)\Phi(P),
\end{equation}
where $P$ is a point in the bulk and $P_{\infty}$ is a boundary point that connects to $P$ by a shortest contour along the radial direction in Fefferman--Graham coordinates. Note that there is an ambiguity in defining the $\Phi$ fields, since they depend on the choice of the contour from $P_{\infty}$ to $P$ \cite{Choi:2018oel,Chen:2019hdv}. We fix the ambiguity of $\Phi$ such that $\Psi$ is contour independent by changing the homotopy class of the contour as we change the definition of $\Phi$. The equation of motion reduces to
\begin{equation}
(\Box - M^2)\Phi=0 .
\end{equation}
In the presence of the gauge field, the thermal periodicity
\begin{equation}
\Psi(t_E+\beta)=e^{+\rmi \mu q } \Psi(t_E),
\end{equation}
follows from the trace cyclicity of the boundary dual charge operator $\mO$ with $[\hat{Q},\mO]=q \mO$, 
\begin{equation}
    \bTr{e^{\rmi \mu \hat{Q}}e^{-\beta \hat{H}}\mO(t_E,x)} = e^{\rmi \mu q}\bTr{e^{\rmi \mu \hat{Q}}e^{-\beta \hat{H}}\mO(t_E-\beta,x)}\; \Rightarrow \mO(t_E+\beta,x) =e^{\rmi \mu q} \mO(t_E,x). 
\end{equation}
This boundary condition induces a twisted boundary condition for \(\Phi\), 
\begin{equation}\label{eq:periodic_BC_Phi}
\Phi(t_E+\beta)=e^{-\rmi\mu q}\,\Phi(t_E),
\end{equation}
which is derived from the flux condition 
where $\Sigma$ is a two-dimensional surface that intersects $\mC$ transversely. 
Consider a mode expansion,
\begin{equation}
\Phi(t_E,z,x) 
= \frac{1}{L}\sum_k\frac{1}{\beta}\sum_{\omega_n}e^{-\rmi\omega_n t_E - {\rmi} k x}\, R_{nk}(z) \Phi_0(\omega_n,k),
\end{equation}
where $\Phi_0(\omega_n,k)$ denotes the source field \cite{Gubser:1998bc,Witten:1998qj}. From the boundary condition \eqref{eq:periodic_BC_Phi}, we can see the Matsubara frequency is shifted:
\begin{equation}
\omega_n
= \frac{2\pi}{\beta}\qty(n+q \frac{\mu}{2\pi}),
\qquad n\in\mathbb{Z}.
\end{equation}
We observe that there is spectral flow for integer $q$ from $\mu=0$ to $\mu=2\pi$. 
The radial component of the wave equation becomes
\begin{equation}
z(1-z) R_{nk}''(z)
+ (1-z) R_{nk}'(z)
+\left(
-\frac{W_n^2}{4z}
-\frac{K^2}{4}
-\frac{M^2}{4(1-z)}
\right) R_{nk}(z)
=0,
\end{equation}
where $W_n:=\frac{\omega_n}{r_0},
\; K:=\frac{k}{r_0}$.
The general solution is a linear combination of
\begin{equation}
\begin{split}
R^{(1)}(z)
&= z^{A}(1-z)^{2 \Delta_-}\,
{}_2F_1(a,b;c;z),\\
R^{(2)}(z)
&= z^{A+1-c}(1-z)^{2\Delta_-}\,
{}_2F_1(a-c+1,b-c;2-c;z),
\end{split}
\end{equation}
where $_2F_1(a,b;c;z)$ is the Gauss hypergeometric function, and the parameters are chosen to be 
\begin{equation}
\begin{split}
& A = \frac{|W_n|}{2} > 0,\quad
{\Delta_\pm}:=\left(1\pm\sqrt{1+M^2}\right),\\
& a+b = 2A+2\Delta_-,\quad
ab = (A+\Delta_-)^2+\frac{K^2}{4},\quad
c = 2A+1 = |W_n|+1 .
\end{split}
\end{equation}
Since $A+1-c=-A<0$, the solution \(R^{(2)}\) is singular at \(z=0\),
and we keep only the \(R^{(1)}\). Using the transformation formula of the hypergeometric function
\begin{equation}
\begin{split}
R^{(1)}(z)
&= z^{A}(1-z)^{2\Delta_+}
\frac{\Gamma(c)\Gamma(a+b-c)}{\Gamma(a)\Gamma(b)}
\,{}_2F_1(c-a,c-b;c-a-b+1;1-z)\\
&\quad
+ z^{A}(1-z)^{2\Delta_-}
\frac{\Gamma(c)\Gamma(c-a-b)}{\Gamma(c-a)\Gamma(c-b)}
\,{}_2F_1(a,b;a+b-c+1;1-z).
\end{split}
\end{equation}
The first term corresponds to the normalizable mode, while the second is a non-normalizable mode. Following the GKPW prescription \cite{Gubser:1998bc,Witten:1998qj}, we can derive the boundary correlator from this solution of the bulk wave equations above. Indeed, the on-shell action of the probe field is given by 
\begin{equation}
    \begin{split}
        I_{\mathrm{scalar}}[\Psi;G,A,\bar{A}] &= -\int_{z=z_\infty} \dd^2 x \sqrt{h}n^A (D_A \Psi)\Psi^\dag= -\int_{z=z_\infty} \dd^2 x \sqrt{h}n^A (\nabla_A \Phi)\Phi^\dag\\
        &= \frac{1}{L}\sum_k\frac{1}{\beta}\sum_{\omega_n}\Phi_0^\dag (k,\omega_n)\qty[-zR_{nk}^*(z)\partial_z R_{nk}(z)]_{z=z_\infty}\Phi_0(k,\omega_n),
    \end{split}
\end{equation}
where $h_{ab}$ is an induced metric on $z_{\infty}=1-\ep$ and we take $\ep\to 0$ after the holographic renormalization. 
Identifying the non-normalizable mode of the $\Psi$ field with $\Phi_0$, as justified by the choice of $P_{\infty}$, we obtain the Green function
\begin{equation}
    G_E^{(\beta)}(\omega_n,k) =\qty[-zR_{nk}^*(z)\partial_z R_{nk}(z)]_{z=z_\infty}.
\end{equation}
Imposing the boundary condition \((1-z)^{-2 \Delta_-}R_{nk}(z)\to 1\) as
\(z\to 1\), and following the holographic renormalization prescription
\cite{Son:2002sd}, we can find correlation functions. For our purpose of proving SWSSB, it is enough to discuss some specific operators, and it turns out to be sufficient to discuss the simplest case \(\Delta_+=2\) (\(\Delta_-=0\)). The Euclidean Green
function then reduces to 
\begin{equation}
G_E^{(\beta)}(\omega_n,k)
= 2\left(\frac{\beta}{2\pi}\right)^2
(\omega_n^2+k^2)
\left[
\psi\!\left(1+\frac{\beta}{4\pi}(\abs{\omega_n}+\rmi k)\right)
+
\psi\!\left(1+\frac{\beta}{4\pi}(\abs{\omega_n}-\rmi k)\right)
\right],
\end{equation}
where $\psi(x)=\Gamma'(x)/\Gamma(x)$ is the digamma function. 
Note that this result was essentially known in \cite{Son:2002sd}, although we are now in the Euclidean signature.

We now evaluate the Fourier transform
\begin{equation}
G_E^{(\beta)}(t_E,x)
= \frac{1}{\beta}
\sum_{\omega_n}\frac{1}{L}\sum_k
e^{-\rmi\omega_n t_E - \rmi k x}\,
G_E^{(\beta)}(\omega_n,k)
\end{equation}
in the large spatial volume limit $L\to \infty$ and we replace $\frac{1}{L}\sum_k \to \int_{-\infty}^\infty \frac{\dd k}{2\pi}$.  Using the fact that digamma function $\psi(z)$ has poles at $z=-n,\;n\in\mathbb{Z}_{\geq0}$ with associated residues $-1$, we can perform the $k$-integrals as 
\begin{equation}
\begin{split}
&\quad \qty(\frac{4\pi}{\beta})^2
\int_{-\infty}^{\infty}\frac{\dd k}{2\pi}\,e^{-\rmi k x}\,
\qty((W-1)^2+\qty(\frac{k\beta}{4\pi})^2)\,
\qty[\psi\qty(W+\rmi \frac{k\beta}{4\pi})+\psi\qty(W-\rmi \frac{k\beta}{4\pi})] \\
&=
\qty(\frac{4\pi}{\beta})^3 \,
e^{-\frac{4\pi}{\beta}W\abs{x}}
\frac{((-2W+3)e^{-\frac{4\pi}{\beta}\abs{x}}+(2W-1))}
{\qty(1-e^{-\frac{4\pi}{\beta}\abs{x}})^3}\, .
\end{split}
\end{equation}
where $W=1+{\beta\abs{\omega_n}}/{(4\pi)}$. The $\omega_n$ summation is a simple geometric series, and we obtain the final expression 
\begin{equation}
\begin{split}\label{eq:holographic_correlation_function}
   &  G_E^{(\beta)}(t_E,x)=  \frac{32\pi}{\beta^2} \\&\times \qty(\frac{w\bar{w}(1+w\bar{w})}{(1-w\bar{w})^3}\qty(\frac{w^\nu}{1-w}+\frac{\bar{w}^{1-\nu}}{1-\bar{w}})+\frac{w\bar{w}}{(1-w\bar{w})^2}\qty(\frac{w^\nu}{(1-w)^2}(\nu+(1-\nu)w)+\frac{\bar{w}^{1-\nu}}{(1-\bar{w})^2}((1-\nu)+\nu \bar{w}))).
\end{split}
\end{equation}
where
 \begin{equation}
     w = e^{-\frac{2\pi}{\beta}(\abs{x}+\rmi t_E)}\;, 
     \quad
     \bar{w}  = e^{-\frac{2\pi}{\beta}(\abs{x}-\rmi t_E)}, 
     \quad
     \nu := \frac{\beta \omega_{n_0}}{2\pi}= 
     \left\{ \frac{q\mu}{2\pi} \right\} .
 \end{equation}
 Here $\{s\}$ denotes the fractional part of $s$,
 and $n_0$ is the smallest integer for which $\omega_n>0$.
 When $\abs{x}\to \infty$ and $t_E=0$, $w\bar{w}\to0$, thus we obtain 
 \begin{equation}
     G_E^{(\beta)}(0,x) \to 0.
 \end{equation}
 This means that there is no long-range order for the usual correlator and there is no weak symmetry breaking. Moreover, we have 
 \begin{equation}
    G_E^{(\beta)}\qty(\frac{\beta}{2}, 0) = \frac{1}{2\pi} \left( \frac{4\pi}{\beta} \right)^2 e^{-\rmi\nu\pi} (1-2\nu)\left(1+\frac{2\nu}{3}-\frac{2\nu^2}{3}\right).
\end{equation}
Note that this expression is invariant under the spectral flow.
Let us perform the Fourier transformation over $\mu$. 
Finally, with large $N$ factorization, we obtain the formula of R\'{e}nyi-2 correlators in terms of Green functions, 
 \begin{align}
   \lim_{\abs{x-y}\to \infty}  R^{(2)}(x,y)
    & = 
    \lim_{\abs{x-y}\to \infty}
    \frac{
        \int_0^{2\pi}\frac{\dd \mu}{2\pi}\, 
        e^{- \rmi \mu Q}
        \bTr{
            e^{\rmi \mu \hat{Q}} 
            e^{-2\beta \hat{H}} 
            \hat{\mO}(x,\beta)
            {\hat{\mO}}(y,\beta)^\dag 
            {\hat{\mO}}(y,0)
            {\hat{\mO}}(x,0)^\dag
        }
    }{
        \int_{0}^{2\pi} \frac{\dd \mu}{2\pi}\; 
        e^{-\rmi \mu {Q}}
        \bTr{
            e^{ \rmi \mu \hat{Q}}
            e^{-2\beta \hat{H}}
        }
    }\nonumber
     \\
    &=
    \frac{
        \int_{0}^{2\pi} \frac{\dd \mu}{2\pi}\;
        e^{-\rmi \mu Q} \,
        \bTr{
            e^{\rmi \mu \hat{Q}}
            e^{-2\beta \hat{H}}}
            \abs{G_E^{(2\beta)}(\beta,0)}^2
        }
    {
        \int_{0}^{2\pi} \frac{\dd \mu}{2\pi}\; 
        e^{-\rmi \mu Q} \,
        \bTr{
            e^{\rmi \mu \hat{Q}}
            e^{-2\beta \hat{H}}
            }
    }\\&=\frac{2}{\pi}\qty(\frac{4\pi}{\beta})^4 \qty(\frac{359}{945}+\sum_{l\geq 1}f_l \cosh{\qty(\frac{\pi \beta q Q}{\kCS L}l)}\exp\qty(-\frac{\pi \beta q^2}{2\kCS L}l^2))
    \geq \frac{2}{\pi}\qty(\frac{4\pi}{\beta})^4  \frac{359}{945}=\LandauO{L^0}.
\label{eq:result}
 \end{align}
where
\begin{equation}
 \begin{split}
       f(\{s\}) &= (1-2\{s\})^2\qty(1+\frac{2\{s\}}{3}-\frac{2\{s\}^2}{3})^2,\;f(s)=\sum_{l\in\mathbb{Z}}f_l e^{2\pi \rmi l s},\; f_l = \int_0^1    \dd s f(s)  e^{-2\pi \rmi l s},\\
        f_0 &= \frac{359}{945},\; 
        f_l=\frac{2 \left(2 \pi ^4 l^4+9 \pi ^2 l^2+30\right)}{3 \pi ^6 l^6},\;l\neq 0.
\end{split}
\end{equation}
We can also consider the Wightman correlator (Rényi-1 correlator) \cite{Liu:2024mme,Weinstein:2024fug}
\begin{equation}
C^W(x,y)=\bTr{\sqrt{\rho}\mO(x)\mO(y)^\dag\sqrt{\rho} \mO(y)\mO(x)^\dag} = \frac{\bTr{\hat{\Pi}_Q e^{-\beta \hat{H}}\mO\qty(x,\frac{\beta}{2})\mO\qty(y,\frac{\beta}{2})^\dag\mO(y)\mO(x)^\dag}}{\bTr{\hat{\Pi}_Q e^{-\beta \hat{H}}}} .
\end{equation}
A computation parallel to the above shows its lower bound in the large distance limit:
\begin{align}
  \lim_{\abs{x-y}\to \infty} C^W(x,y)&= \lim_{\abs{x-y}\to \infty}
    \frac{
        \int_0^{2\pi}\frac{\dd \mu}{2\pi}\, 
        e^{- \rmi \mu Q}
        \bTr{
            e^{\rmi \mu \hat{Q}} 
            e^{-\beta \hat{H}} 
            \hat{\mO}\qty(x,\frac{\beta}{2})
            {\hat{\mO}}\qty(y,\frac{\beta}{2})^\dag
            {\hat{\mO}}(y,0)
            {\hat{\mO}}(x,0)^\dag
        }
    }{
        \int_{0}^{2\pi} \frac{\dd \mu}{2\pi}\; 
        e^{-\rmi \mu {Q}}
        \bTr{
            e^{ \rmi \mu \hat{Q}}
            e^{-\beta \hat{H}}
        }
    }\nonumber
     \\
    &=
    \frac{
        \int_{0}^{2\pi} \frac{\dd \mu}{2\pi}\;
        e^{-\rmi \mu Q} \,
        \bTr{
            e^{\rmi \mu \hat{Q}}
            e^{-\beta \hat{H}}}
            \abs{G_E^{(\beta)}\qty(\frac{\beta}{2},0)}^2
        }
    {
        \int_{0}^{2\pi} \frac{\dd \mu}{2\pi}\; 
        e^{-\rmi \mu Q} \,
        \bTr{
            e^{\rmi \mu \hat{Q}}
            e^{-\beta \hat{H}}
            }
    }\geq \frac{2}{\pi}\qty(\frac{8\pi}{\beta})^4  \frac{359}{945}=\LandauO{L^0}.
    \label{eq:result2}
\end{align}
\end{document}